\documentclass[prx,twocolumn,superscriptaddress]{revtex4-2}

\usepackage{graphicx,float}
\usepackage{braket}
\usepackage{siunitx}
\usepackage{appendix}
\usepackage{amssymb}
\usepackage{mathrsfs}
\usepackage{appendix}
\usepackage{slashed}
\usepackage{physics}
\usepackage{mathtools}
\usepackage{amsmath}
\usepackage{bbold}

\newcolumntype{Y}{>{\centering\arraybackslash}X}

\begin{document}

\title{Efficient simulation of ultrafast quantum nonlinear optics with matrix product states}

\author{Ryotatsu Yanagimoto}
\thanks{Email: ryotatsu@stanford.edu}
\affiliation{E.\,L.\,Ginzton Laboratory, Stanford University, Stanford, California 94305, USA}

\author{Edwin Ng}
\affiliation{E.\,L.\,Ginzton Laboratory, Stanford University, Stanford, California 94305, USA}

\author{Logan G. Wright}
\affiliation{NTT Physics and Informatics Laboratories, NTT Research, Inc., 1950 University Ave., East Palo Alto, California 94303, USA}
\affiliation{School of Applied and Engineering Physics, Cornell University, Ithaca, New York 14853, USA}

\author{Tatsuhiro Onodera}
\affiliation{NTT Physics and Informatics Laboratories, NTT Research, Inc., 1950 University Ave., East Palo Alto, California 94303, USA}
\affiliation{School of Applied and Engineering Physics, Cornell University, Ithaca, New York 14853, USA}

\author{Hideo Mabuchi}
\affiliation{E.\,L.\,Ginzton Laboratory, Stanford University, Stanford, California 94305, USA}

\date{\today}

\begin{abstract}
Ultra-short pulses propagating in nonlinear nanophotonic waveguides can simultaneously leverage both temporal and spatial field confinement, promising a route towards single-photon nonlinearities in an all-photonic platform. In this multimode quantum regime, however, faithful numerical simulations of pulse dynamics na{\"i}vely require a representation of the state in an exponentially large Hilbert space. Here, we employ a time-domain, matrix product state (MPS) representation to enable efficient simulations by exploiting the entanglement structure of the system. In order to extract physical insight from these simulations, we develop an algorithm to unravel the MPS quantum state into constituent temporal supermodes, enabling, e.g., access to the phase-space portraits of arbitrary pulse waveforms. As a demonstration, we perform exact numerical simulations of a Kerr soliton in the quantum regime. We observe the development of non-classical Wigner-function negativity in the solitonic mode as well as quantum corrections to the semiclassical dynamics of the pulse. A similar analysis of $\chi^{(2)}$ simultons reveals a unique entanglement structure between the fundamental and second harmonic. Our approach is also readily compatible with quantum trajectory theory, allowing full quantum treatment of propagation loss and decoherence. We expect this work to establish the MPS technique as part of a unified engineering framework for the emerging field of broadband quantum photonics.
\end{abstract}

\maketitle

\section{Introduction}
The ability to manipulate photon-photon interactions at the quantum level holds the key to lifting conventional classical limits in a wide range of photonic technologies and applications~\cite{LIGO2013, Tsang2016, Gisin2007, Obrien2009, Chang2014, Zhong2020}. Recent efforts in the field of nonlinear nanophotonics have resulted in the development of ultra-low-loss and highly efficient platforms for nonlinear optics~\cite{Zhang2017, Wang2018}, with experimental numbers coming remarkably close to bridging the long-standing gap between classical optics and the ``strong interaction regime'' of quantum optics with single-photon-level nonlinearity~\cite{Lu2020, Placke2020, Ramelow2019, Hueck2020, Bruch2019}. In particular, advances in dispersion engineering on these platforms enable ultra-short-pulse operation~\cite{Jankowski2020, Zhang2012}, where the available peak power further leverages the material nonlinearities by orders of magnitude, bringing the possibility of engineering highly non-classical states of light into the foreseeable future~\cite{Yanagimoto2020}. In the presence of such strong nonlinearities, the quantum nature of individual photons plays a critical role in the physical behavior of these systems~\cite{Birnbaum2005}, i.e., classical mean-field theories and semiclassical approximations are no longer valid in predicting the results of experiments~\cite{Javanainen2016, Yanagimoto2020_fano}. At the same time, controllably harnessing these exotic quantum phenomena for photonic quantum engineering requires the development of unified simulation and modeling methodologies faithful to the hardware level, posing a significant and imminent theoretical and modeling problem. In general, modeling non-classical photon dynamics in a broadband system like a strongly nonlinear propagating pulse is a nontrivial task due to the immense dimension of the Hilbert space that the quantum optical states in general occupy. For instance, when we discretize an optical pulse into $100$ bins, as is conventionally done in classical pulse propagation, but with, say, $<10$ interacting photons in each bin, we would na{\"i}vely need to compute the evolution of a $\sim 10^{100}$-dimensional quantum state vector, which clearly exceeds any available computational resource. Thus, to fully exploit the technological advantages offered by these emerging quantum-optical devices, it is essential to apply sophisticated model reduction techniques to the na{\"i}ve full quantum model to obtain computationally tractable models that nevertheless retain quantum features expected to be important given the hardware specifications.

For modeling the propagation of a quantum pulse, some key features that we can utilize to reduce the complexity of the na{\"ive} quantum state are (1) the one-dimensionality of the optical field, and (2) the locality of the interactions and dynamics. Quantum mechanically, a nonlinear waveguide can be seen as a system of bosons (i.e., photons) confined to a one-dimensional space evolving under a Hamiltonian that mediates only local interactions, and intuitively, the classical envelope of a pulse (i.e., the $g^{(1)}$ correlation function) simply describes how the bosons are spatially distributed~\cite{Drummond2014}. Notably, it is known in the context of many-body physics that such a one-dimensional system of locally interacting particles often exhibits a limited amount of entanglement~\cite{Vidal2004}, indicating that a large portion of the na{\"i}vely large Hilbert space (i.e., the parts representing highly nonlocal entangled states) are not relevant. One technique extensively used in the field of many-body physics to exploit this feature is to cast the quantum state into the form of a matrix product state (MPS)~\cite{Schollwoeck2011, Orus2014}, which yields an efficient state representation for numerical studies~\cite{Muth2010_b, Daley2012}.

MPS has recently been extended to analyze optical systems, such as in the treatment of pulse propagation through a mesoscopic atomic cloud~\cite{Manzoni2017} and atom-coupled chiral waveguides~\cite{Mahmoodian2020}. While these studies have uncovered rich many-body dynamics, the focus has predominantly been on the ``particle'' aspect of the physics, such as analyzing their mean-field distributions and $n$-particle correlation functions. On the other hand, relatively little attention has been paid to the phase-space properties of the quantum state~\cite{Cahill1969} which characterize the ``oscillator'' aspect of the physics related to the continuous-variable (CV) nature of the optical field~\cite{Braunstein2005, Gottesman2001}; in many cases, it is this latter picture that conceptually connects more directly to the perspectives employed both in classical wave optics and CV quantum information. In the CV approach, it is natural to ask about, for example, the reduced density matrix~\cite{Nielsen2000} of a given pulse supermode, which is indispensable for photonic quantum state engineering in the time~\cite{Humphreys2013, Brecht2015, Asavarant2019, Ansari2018} and/or spectral~\cite{Lukens2017, Roslund2014} domains, but which is not straightforward to infer from the $n$-particle correlation functions. Other CV quantities such as the Wigner function and various entanglement measures computed from the reduced density matrix are routinely used to quantify photonic states as resources for quantum information processing~\cite{Chitambar2019, Albarelli2018}.

In this article, we first review the application of MPS and MPS-based time evolution methods to the problem of quantum pulse propagation. In order to appreciate the optical phase-space dynamics of this novel nonlinear quantum regime, we develop a ``demultiplexing'' scheme which can be applied to an MPS representation of a quantum pulse to extract the reduced density matrix in a basis of pulse supermodes~\cite{Brecht2015}. In our scheme, a carefully chosen sequence of one- and two-mode (i.e., local) linear operations (phaseshifters and beamsplitters) are used to overcome the problem of manipulating nonlocal supermodes---which are not naturally accessible in the MPS framework---by mapping them onto local bins. As a demonstration, we analyze the quantum propagation of a pulse initialized in a coherent-state Kerr soliton of a 1D $\chi^{(3)}$-nonlinear waveguide~\cite{Agrawal2019}, using the MPS-based approach of time-evolving block decimation (TEBD)~\cite{Vidal2003, Vidal2004, Garcia-Ripoll2006} which is compatible with standard quantum-optical treatments of linear loss~\cite{Wiseman2009}. We show that the Wigner function of the state contained in the canonical sech-pulse supermode can exhibit a considerable amount of negativity with an enhancement correlated with the peak intensity. We highlight features of the pulse dynamics that indicate departures from established models, such as broadband corrections to the conventional time-dependent Hartree-Fock approximation for fundamental Kerr solitons~\cite{Haus1989,Wright1991}, and, for higher-order solitons~\cite{Kivshar2003}, stark qualitative deviations from classical breather dynamics. Then, we extend our analysis to the quantum propagation of a pulse initialized as a coherent-state simulton~\cite{Werner1993}, a quadratic soliton of a 1D $\chi^{(2)}$-nonlinear waveguide composed of two co-propagating pulses at the fundamental harmonic (FH) and second harmonic (SH). We compute a two-mode reduced density matrix for the FH and SH pulse supermodes and reveal their entanglement structure. As a result, we find that non-classicality in the quantum simulton primarily accumulates in a hybrid supermode consisting of a specific linear combination of the FH and SH supermodes. The numerical techniques highlighted in these examples can be generalized to address various questions about quantum pulse propagation expected to arise with the advent of strongly-interacting broadband quantum photonics.

\section{Matrix Product States for Quantum Optical Pulse Propagation}
\label{sec:mps}
In this section, we introduce basic concepts of matrix product states (MPS), together with methods to implement time evolutions. While we base our discussions on quantum pulse propagation on a 1D $\chi^{(3)}$-nonlinear waveguide to keep discussions concrete, the basic concepts introduced in this section is general and can be applied to a broader class of systems. We extend our analysis to $\chi^{(2)}$ nonlinear waveguides in Sec.~\ref{sec:simulton}.

The Hamiltonian of a 1D $\chi^{(3)}$-nonlinear waveguide in the moving frame takes a form~\cite{Drummond2014}
\begin{equation}
    \label{eq:liebliniger}
    \hat{H}=-\frac{1}{2}\int \mathrm{d}z\left(\hat{\phi}_z^\dagger\partial_z^2\hat{\phi}_z+\hat{\phi}_z^{\dagger 2}\hat{\phi}^2_z\right),
\end{equation}
where photon field annihilation operators fulfill commutation relationships $[\hat{\phi}_z,\hat{\phi}_{z'}^\dagger]=\delta(z-z')$. The mean-field (i.e., c-number) equation under \eqref{eq:liebliniger} takes a well-known form of nonlinear Schr\"odinger equation (NLSE)~\cite{Agrawal2019}
\begin{align}
    \label{eq:nlse}
    \mathrm{i}\partial_t\phi_z=-\frac{1}{2}\partial_z^2\phi_z-|\phi_z|^2\phi_z,
\end{align}
where time $t$ and space $z$ have been normalized. In the context of many-body physics, \eqref{eq:liebliniger} is referred to as Lieb-Liniger Hamiltonian, describing bosons (e.g., photons) with point-like interactions ~\cite{Lieb1963,Muth2010_b}. 

\begin{figure}[tb]
    \centering
      \includegraphics[width=0.43\textwidth]{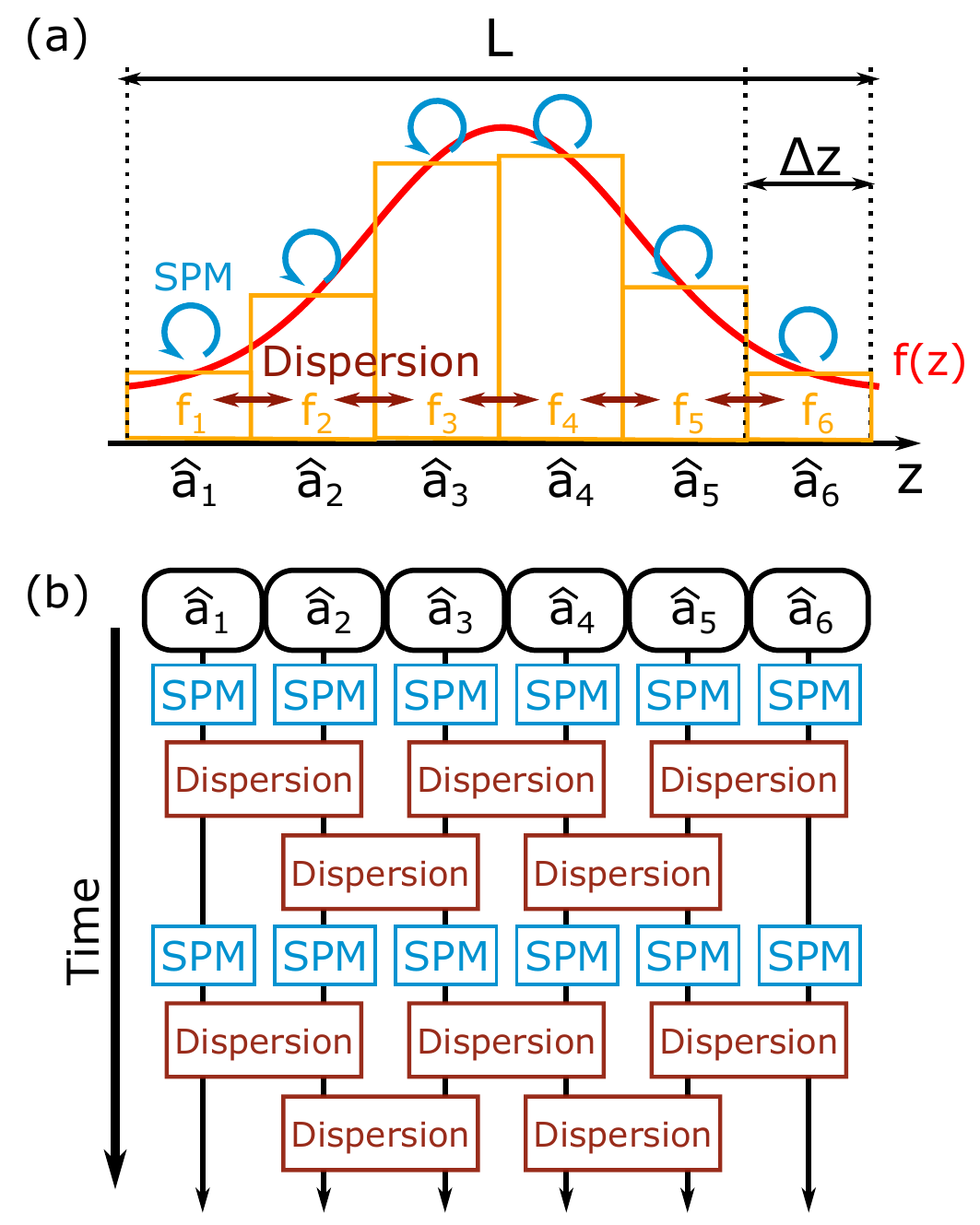}
      \caption{(a) Finite space interval $-L/2\leq z\leq L/2$ is discretized to $n$ spatial bins with size $\Delta z=L/N$. Pulse envelope function $f(z)$ (red lines) is discretized to form a vector $\boldsymbol{f}=(f_1,\dots,f_N)^\intercal$ (orange boxes). Self-phase modulation (SPM) acts locally on discretized photon field (blue arrows), while dispersion mediates interactions between neighboring bins (brown arrows). (b) An example of how time evolution under the Hamiltonian \eqref{eq:bosehubbard} can be decomposed into one-mode SPM operations (blue boxes) and two-mode operations corresponding to dispersions (brown boxes). Spatial bins are labeled by corresponding annihilation operators for the purpose of illustration.
    }
      \label{fig:chi3}
  \end{figure}

While the continuum coordinate $z$ is convenient for analytic studies, it is often easier to work in a discretized coordinate for numerical evaluations. More importantly, discretization of the field allows us to encode system states on an MPS. We consider a finite space interval $-L/2\leq z\leq L/2$ and discretize it into $N$ spatial bins with size $\Delta z=L/N$. As shown in Fig.~\ref{fig:chi3} (a), we assign a photon annihilation operator $\hat{a}_m$ to the $m$th the spatial bin for $m\in\{1,2,\dots,N\}$. When $N$ is large enough, \eqref{eq:liebliniger} can be approximated by a discretized Bose-Hubbard Hamiltonian of the form~\cite{Muth2010,Muth2010_c}
\begin{align}
    \label{eq:bosehubbard}
    \begin{split}
    \hat{H}=&\sum_m\left[-\frac{1}{2\Delta z^2}\left(\hat{a}_m^\dagger\hat{a}_{m+1}+\mathrm{H.c.}\right)\right.\\
    &\qquad\left.+\frac{1}{\Delta z^2}\hat{a}_m^\dagger\hat{a}_m-\frac{1}{2\Delta z}\hat{a}_m^{\dagger 2}\hat{a}_m^2\right].
    \end{split}
\end{align}
Similarly, as shown in the figure, a pulse waveform in the continuous coordinate $f(z)$ is discretized as an $N$-dimensional vector $\boldsymbol{f}=(f_1,\dots,f_N)^\intercal$.

A generic system state can be written as
\begin{align}
    \label{eq:genericstate}
    \ket{\Psi}=\sum_{\boldsymbol{i}}c_{\boldsymbol{i}}\ket{\boldsymbol{i}},
\end{align}
where $\boldsymbol{i}=(i_1,i_2,\dots,i_N)^\intercal$ and $\ket{\boldsymbol{i}}=\ket{i_1}\otimes\ket{i_2}\cdots \ket{i_N}$. Each local Hilbert space is spanned by Fock states $\ket{i_m}~(i_m\geq 0)$. Generally, keeping and updating $c_{\boldsymbol{i}}$ require computational resources that grow exponentially with respect to $N$. On the other hand, depending on the specific properties of the Hamiltonian, such as locality, much of the states in the entire Hilbert space might not be populated and thus could be excluded. To this end, matrix product states (MPS) allows an effective representation of quantum states given the amount of entanglement is limited and local. 

Using an MPS representation with bond dimension $\chi$, \eqref{eq:genericstate} can be approximated as~\cite{Vidal2003,Vidal2004}
\begin{align}
    \label{eq:mpsapprox}
    c_{\boldsymbol{i}}=\sum_{\alpha_1,\dots,\alpha_{n-1}}\Gamma^{[1]i_1}_{1\alpha_1}\lambda^{[1]}_{\alpha_1}\Gamma^{[2]i_2}_{\alpha_1\alpha_2}\lambda^{[2]}_{\alpha_2}\cdots\lambda^{[N-1]}_{\alpha_{N-1}}\Gamma^{[N]i_N}_{\alpha_{N-1}1}~,
\end{align}
where $\Gamma^{[m]}$ is a rank-3 tensor, $\lambda^{[m]}$ is a vector, and each $\alpha_m$ runs through $1$ to $\chi$. Intuitively, \eqref{eq:mpsapprox} is a decomposition of a rank-$N$ tensor $c_{\boldsymbol{i}}$ into a product of low-rank tensors. For instance, a coherent pulse with a normalized envelope $\boldsymbol{f}$ with $\sum_m|f_m|^2=1$ takes a form
\begin{align}
    \label{eq:coherent}
    &\Gamma^{[m]i_m}_{11}=e^{-|f_m^*\alpha|^2/2}\frac{(f_m^*\alpha)^{i_m}}{\sqrt{i_m!}},&\lambda_{1}^{[m]}=1,
\end{align}
where all the other tensor elements are zero. Physically, \eqref{eq:coherent} is obtained by displacing a supermode $\hat{A}=\sum_mf_m\hat{a}_m$ by $\alpha$ from the vacuum. Notice that parameters needed for \eqref{eq:mpsapprox} has a favorable polynomial scaling of $\mathcal{O}(\chi^2N)$. The bond dimension $\chi$ is related to the maximum amount of entanglement that \eqref{eq:mpsapprox} can support, and larger $\chi$ is needed to describe $\ket{\Psi}$ with longer-range entanglement. Notably, it is heuristically known that entanglement in 1D quantum many-body systems is often limited~\cite{Vidal2004}, making MPS an ideal representation to study quantum pulse propagation.

Similarly, a generic operator 
\begin{align}
    \hat{O}=\sum_{\boldsymbol{i}\boldsymbol{i}'}O_{\boldsymbol{i}\boldsymbol{i}'}\ketbra{\boldsymbol{i}}{\boldsymbol{i}'}
\end{align}
can also be expressed in the form of a matrix product as
\begin{align}
    \label{eq:mpo}
    O_{\boldsymbol{i}\boldsymbol{i}'}=\sum_{\alpha_1,\dots,\alpha_{N-1}}O^{[1]i_1i_1'}_{1\alpha_{1}}O^{[2]i_2i_2'}_{\alpha_1\alpha_2}\cdots O^{[N]i_Ni_N'}_{\alpha_{N-1}1}
\end{align}
where $O^{[m]}$ is a rank-4 tensor. Operators expressed in the form of \eqref{eq:mpo} are referred to as matrix product operators (MPO), and their expectation values with respect to MPS are computed via tensor contractions~\cite{Schollwoeck2011}. Just in the same way as $\chi$ for an MPS is determined by how entangled the state is, bond dimension of an MPO is determined by how non-local the operator $\hat{O}$ is.

In his seminal paper~\cite{Vidal2003}, Vidal introduced an algorithm, which is often referred to as time-evolving block decimation (TEBD), to efficiently simulate the time evolution of an MPS. Multiple open-source packages have been developed for TEBD~\cite{Jaschke2018, Bauer2011} to this date. TEBD utilizes the fact that updating an MPS for local one-mode or two-mode unitary operations can be done efficiently. 
While this process involves a truncation of minor singular-value components, truncation error can in principle be arbitrary small by taking large enough bond dimension $\chi$. For instance, as shown in Fig.~\ref{fig:chi3}(b), dynamics under \eqref{eq:bosehubbard} is simulated by Trotter decomposing a short-time evolution $\hat{U}=e^{-\mathrm{i}\hat{H}\delta t}$ into a sequence of one-mode and two-mode operations
\begin{align}
    \label{eq:trotter}
    \hat{U}\approx (\hat{D}_{2}\hat{D}_{4}\hat{D}_6\cdots)(\hat{D}_{1}\hat{D}_{3}\hat{D}_5\cdots)(\hat{S}_1\hat{S}_2\hat{S}_3\cdots),
\end{align}
where $\hat{S}_m=\exp(\mathrm{i}\delta t\hat{a}_m^{\dagger 2}\hat{a}_m^2/2\Delta z)$ implements Kerr self-phase modulation (SPM) on the $m$th bin, and 
\begin{align}
\begin{split}
    \hat{D}_m=\exp&\left[\frac{\mathrm{i}\delta t}{2\Delta z^2}(\hat{a}_m^\dagger\hat{a}_{m+1}+\hat{a}_m\hat{a}_{m+1}^\dagger\right.\\
    &\qquad\qquad\left.-\hat{a}_m^\dagger\hat{a}_m-\hat{a}_{m+1}^\dagger\hat{a}_{m+1})\right]
    \end{split}
\end{align}
represents hopping interactions between $m$th and $(m+1)$th bins due to quadratic energy dispersions. Note that operations in each parentheses of \eqref{eq:trotter} commute, and thus, they can be implemented in parallel numerically. Additionally, higher-order decomposition methods can reduce the Trotter discretization errors~\cite{Sornborger1999}. For comprehensive overview on time-evolution methods for MPS, we lead readers to Ref.~\cite{Garcia-Ripoll2006}.

It is worth mentioning that we can readily include dissipations to the simulation, e.g., by the Monte-Carlo wavefunction method (MCWF)~\cite{Wiseman2009}. The capability of including loss is particularly important for optical simulations, not only because realistic optical systems often have a non-negligible amount of loss, but also because dissipation plays critical roles in a host of emergent phenomena, such as dissipative Kerr solitons~\cite{Kippenberg2018} and the physics of $\mathcal{PT}$-symmetric systems~\cite{El-Ganainy2007, Alexeeva2012}. 

\section{Demultiplexing supermodes from an MPS}
\label{sec:demultiplexing}
After a numerical simulation of a quantum pulse propagation using MPS, we wish to calculate various physical properties of the resultant quantum state $\ket{\Psi(t)}$. For instance, the two photon correlation function $g^{(2)}(\ell,m)=\langle\hat{a}_\ell^\dagger\hat{a}_m^\dagger\hat{a}_m\hat{a}_\ell\rangle/\langle\hat{a}_\ell^\dagger\hat{a}_\ell\rangle\langle\hat{a}_m^\dagger\hat{a}_m\rangle$ can be calculated by expressing $\hat{a}_\ell^\dagger\hat{a}_m^\dagger\hat{a}_m\hat{a}_\ell$ and $\hat{a}_j^\dagger\hat{a}_j~(j=\ell,m)$ as MPOs and computing their expectation values. This procedure can be extended to compute $n$-particle correlation function as well. Generically, physical quantities associated to MPOs with larger bond dimension are more expensive to compute. 

In the context of quantum engineering and information, it is of great importance to know quantum states populating certain supermodes of interest. To be more precise, let us consider $s\ll N$ supermodes, where the $r$th pulse supermode~\cite{Brecht2015} is defined as
\begin{align}
    \hat{A}^{(r)}=\sum_{m=1}^Nf_m^{(r)}\hat{a}_m
\end{align} 
for an arbitrary set of orthonormal vectors $\{\boldsymbol{f}^{(1)}, \dots, \boldsymbol{f}^{(s)}\}$. We denote the Hilbert space for these supermodes as $\mathcal{S}=\mathcal{S}_1\otimes\mathcal{S}_2\otimes\dots\mathcal{S}_s$, where $\mathcal{S}_r$ is the space for the $r$th supermode, and the Hilbert space for the rest of the system is denoted as $\mathcal{E}$. Here, we are interested in a reduced density matrix
\begin{align}
    \hat{\rho}_\mathcal{S}=\mathrm{Tr}_\mathcal{E}\ketbra{\Psi}{\Psi}=\sum_{\boldsymbol{j}\boldsymbol{j}'}\rho_{\boldsymbol{j}\boldsymbol{j}'}\ket{\boldsymbol{j}}_\mathcal{S}\bra{\boldsymbol{j}'}_\mathcal{S},
\end{align}
where $\boldsymbol{j}=(j_1,\dots,j_s)^\intercal$, $\ket{\boldsymbol{j}}_\mathcal{S}=\ket{j_1}_{\mathcal{S}_1}\otimes \ket{j_2}_{\mathcal{S}_2}\otimes \cdots \otimes \ket{j_s}_{\mathcal{S}_s}$, and similarly for $\ket{\boldsymbol{j}'}_\mathcal{S}$. Once we obtain $\hat{\rho}_\mathcal{S}$, whose dimension is much smaller than the original Hilbert space, we can study detailed properties of the state, such as Wigner function and entanglement among supermodes with standard techniques for continuous variable systems~\cite{Braunstein2005}.

To obtain $\hat{\rho}_\mathcal{S}$, we ideally would like to have a low-rank MPO representation of the operators
\begin{align}
    \hat{\mu}_{\boldsymbol{j}\boldsymbol{j}'}= \ket{\boldsymbol{j}'}_\mathcal{S}\bra{\boldsymbol{j}}_\mathcal{S}\otimes \hat{\mathbb{1}}_\mathcal{E},
\end{align}
where $\hat{\mathbb{1}}_\mathcal{E}$ is an identity operator on $\mathcal{E}$. The expectation values of these operators would directly give the matrix elements of $\hat{\rho}_\mathcal{S}$ via $(\hat{\rho}_\mathcal{S})_{\boldsymbol{j}\boldsymbol{j}'}=\expval{\hat{\mu}_{\boldsymbol{j}\boldsymbol{j}'}}{\Psi}$, but the highly non-local structure of $\hat{\mu}_{\boldsymbol{j}\boldsymbol{j}'}$ makes it nontrivial to find such a low-rank MPO expression.

\begin{figure}[t]
    \centering
      \includegraphics[width=0.43\textwidth]{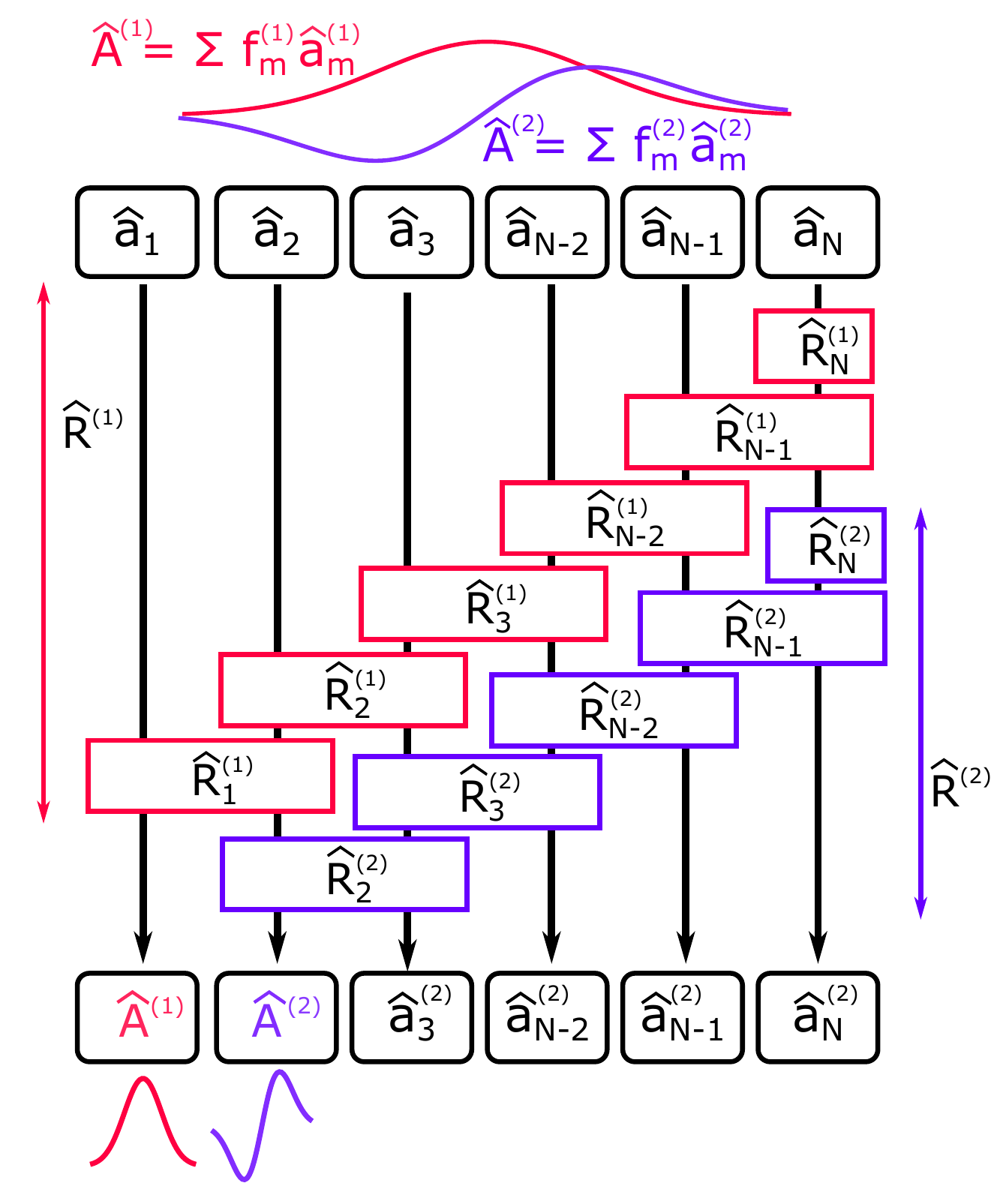}
      \caption{Illustration of our supermode demultiplexing scheme. Application of a unitary transformation $\hat{R}^{(r)}$, which is composed of one-mode and two-mode gates $\{\hat{R}^{(r)}_r,\dots,\hat{R}^{(r)}_N\}$, demultiplexes $r$th supermode $\hat{A}^{(r)}=\sum_m f^{(r)}_m\hat{a}_m$ to the $r$th (local) spatial bin. We show operations up to $r=2$, where local annihilation operators of the first and second bins are transformed to $\hat{A}^{(1)}$ and $\hat{A}^{(2)}$, respectively. For the purpose of illustration, spatial bins are labeled by associated annihilation operators.} 
      \label{fig:generic}
  \end{figure}

In the following, as a main result of this research, we describe a procedure to efficiently calculate the reduced density matrix $\hat{\rho}_\mathcal{S}$ of an MPS $\ket{\Psi}$ for an arbitrary set of supermodes comprising $\mathcal S$. As shown schematically in Fig.~\ref{fig:generic}, our approach is based on constructing a linear unitary operation $\hat{V}$ which ``demultiplexes'' the $s$ (nonlocal) supermodes into the leftmost $s$ (local) spatial bins. In other words, by computing $\ket{\Phi}=\hat{V}\ket{\Psi}$, we would like to calculate the matrix elements of $\hat{\rho}_\mathcal{S}$ via
\begin{align}
    \label{eq:matelements}
    (\hat{\rho}_\mathcal{S})_{\boldsymbol{j}\boldsymbol{j}'}=\expval{\hat{\pi}_{\boldsymbol{j}\boldsymbol{j}'}}{\Phi}
\end{align}
where the operators $\hat \pi_{\boldsymbol{j}\boldsymbol{j'}}$ have a local form
\begin{align}
    \hat{\pi}_{\boldsymbol{j}\boldsymbol{j}'}=\left(\bigotimes_{m=1}^{s} \ketbra{j_m'}{j_m}\right) \otimes \left(\bigotimes_{m=s+1}^{N}\hat{\mathbb{1}}_m \right),
\end{align}
where $\hat{\mathbb{1}}_m$ is an identity operator on the $m$th spatial bin. Explicitly, the MPO representation of $\hat{\pi}_{\boldsymbol{j}\boldsymbol{j}'}$ is
\begin{align}
    \label{eq:pimpo}
    O^{[m]i_mi_m'}_{11}=\left\{\begin{array}{ll}
      \delta_{i_mj_{m}'}\delta_{i_m'j_{m}}&m\leq s\\
        \delta_{i_mi_m'}&s<m
    \end{array}\right.,
\end{align}
with all other tensor elements zero; because \eqref{eq:pimpo} only requires a bond dimension of $1$, its expectation value can be efficiently computed. Then the demultiplexing problem posed by \eqref{eq:matelements} is to obtain
\begin{align}
    \label{eq:condition}
      \hat{V}^{\dagger}\hat{\pi}_{\boldsymbol{j}\boldsymbol{j}'}\hat{V}=\hat{\mu}_{\boldsymbol{j}\boldsymbol{j}'}~.
\end{align}
While solutions to \eqref{eq:condition} are not generally unique, we choose to construct $\hat{V}$ as a product of local one-mode and two-mode linear operations to allow for the computation of $\ket{\Phi}=\hat{V}\ket{\Psi}$ using the standard MPS operations also used in TEBD. Such construction of a linear unitary gate with one-mode and two-mode operations may be seen as a family of general multiport linear interferometers~\cite{Clements2016, Reck1994}, where each ``port'' is a discretized spatial bin of an MPS in our setup.

Our scheme to construct $\hat{V}$ works in an iterative manner with iteration index $r\in\{1,2,\dots,s\}$. In the $r$th iteration, we construct a transformation $\hat{R}^{(r)}$ that demultiplexes the $r$th supermode into the $r$th spatial bin. Note that, as a consequence, the construction of $\hat R^{(r)}$ is dependent on the partial operations
\begin{align}
    \hat{V}^{(r-1)}=\hat{R}^{(r-1)}\hat{R}^{(r-2)}\cdots \hat{R}^{(1)}
\end{align}
which have already been applied. If the previous $(r-1)$ iterations were valid, then we can suppose $\hat{V}^{(r-1)}$ implements the transformations $\hat{a}_m \mapsto \hat{a}^{(r-1)}_m=\hat{V}^{(r-1)\dagger}\hat{a}_m\hat{V}^{(r-1)}$, where
\begin{align}
    \label{eq:lineartrans}
    \hat{a}^{(r-1)}_m=\left\{\begin{array}{ll}
      \hat{A}^{(m)}&m\leq r-1\\
      \sum_{\ell=m-r+1}^N c_{m\ell}^{(r-1)}\hat{a}_\ell&m\geq r
    \end{array}\right..
\end{align}
Here, the first line of \eqref{eq:lineartrans} means the previous $(r-1)$ iterations have successfully demultiplexed supermodes $1$ through $(r-1)$ into the leftmost $(r-1)$ spatial bins. The second line indicates that for spatial bins with index $m\geq r$, the effect of the transformation $\hat{V}^{(r-1)}$ has been to only ``mix in'' components from bins $m-r+1$ up to $N$; more precisely, for $m \geq r$, $\hat{a}^{(r-1)}_m$ is independent of any $\hat{a}_\ell$ such that $\ell\leq m-r$. \eqref{eq:lineartrans} is fulfilled for the base case of $\hat{V}^{(0)}=\hat{\mathbb{1}}$ and $c_{m\ell}^{(0)}=\delta_{m\ell}$ at $r = 0$, so we only need to ensure that our construction of $\hat{R}^{(r)}$ preserves \eqref{eq:lineartrans} for any $r > 0$. At the end of the final $s$th iteration, we should then find that $\hat{a}^{(s)}_m=\hat{A}^{(s)}_m$ for $1 \leq m \leq s$, which satisfies our goal of \eqref{eq:condition} and allows us to take $\hat{V}=\hat{V}^{(s)}$.

We parametrize $\hat{R}^{(r)}$ with two sets of angles $\{\theta_{r}^{(r)}, \dots, \theta_{N}^{(r)}\}$ and $\{\varphi_{r}^{(r)}, \dots, \varphi_{N-1}^{(r)}\}$ (i.e., the $r$th iteration requires $2(N-r)+1$ parameters). For $m < N$, these parameters are taken to define a set of phaseshifter/beamsplitter operations
\begin{subequations} \label{eq:mappings}
\begin{align}
    &\hat{R}^{(r)}_m=\exp\left[\mathrm{i}\theta_m^{(r)}(\hat{a}_{m}^\dagger\hat{a}_m-\hat{a}_{m+1}^\dagger\hat{a}_{m+1})\right]\\
    &\times \exp\left[\left(\frac{\pi}{2}-\varphi_m^{(r)}\right)\left(e^{-\mathrm{i}\theta_m^{(r)}}\hat{a}_m^\dagger\hat{a}_{m+1}-e^{\mathrm{i}\theta^{(r)}_m}\hat{a}_m\hat{a}_{m+1}^\dagger\right)\right],\nonumber
\end{align}
which implement the operator  transformations~\cite{Olivares2012}
\begin{align}
    \hat{R}^{(r)\dagger}_m\hat{a}_m\hat{R}^{(r)}_m&=e^{\mathrm{i}\theta_m^{(r)}}\sin\varphi_m^{(r)}\hat{a}_m+\cos\varphi_m^{(r)}\hat{a}_{m+1}\\
    \hat{R}^{(r)\dagger}_m\hat{a}_{m+1}\hat{R}^{(r)}_m&=e^{-\mathrm{i}\theta_m^{(r)}}\sin\varphi_m^{(r)}\hat{a}_{m+1}-\cos\varphi_m^{(r)}\hat{a}_{m}.\nonumber
\end{align}
For $m = N$, we fix $\hat{R}_N^{(r)}=e^{\mathrm{i}\theta_{N}^{(r)}\hat{a}_N^\dagger\hat{a}_N}$. We then construct $\hat R^{(r)}$ by cascading these one- and two-mode operations according to
\begin{align}
    \hat{R}^{(r)}=\hat{R}^{(r)}_{r}\hat{R}^{(r)}_{r+1}\cdots\hat{R}_{N-1}^{(r)}\hat{R}_N^{(r)}.
\end{align}
\end{subequations}

Because $\hat{R}^{(r)}$ should demultiplex the $r$th supermode into the $r$th spatial bin, we require
\begin{align}
    \label{eq:reference}
    \hat{a}_{r}^{(r)}=\hat{A}^{(r)}=\sum_{\ell=1}^Nf^{(r)}_\ell\hat{a}_\ell.
\end{align}
On the other hand, based on \eqref{eq:lineartrans} and \eqref{eq:mappings}, we have
\begin{align}
\label{eq:trans}
\begin{split}
    \hat{a}_{r}^{(r)}&=\sum_{m=r}^{N}g_m^{(r)}\hat{a}_m^{(r-1)}\\
    &=\sum_{\ell=1}^N\sum_{m=r}^{\min(r-1+\ell,N)}g_m^{(r)}c_{m\ell}^{(r-1)}\hat{a}_\ell,
    \end{split}
\end{align}
where $g_m^{(r)}=e^{\mathrm{i}\theta^{(r)}_{m}}\sin\varphi^{(r)}_{m}\prod^{m-1}_{k=r}\cos\varphi_k^{(r)}$. By equating \eqref{eq:reference} and \eqref{eq:trans}, we can solve for the angles via
\begin{subequations}
\begin{align}
    \label{eq:angles}
    &\theta_{m}^{(r)}=\arg \mathcal{I}_{m}&\varphi_{m}^{(r)}=\sin^{-1} |\mathcal{I}_{m}|,
\end{align}
where
\begin{align}
  \mathcal{I}_{m}=\frac{f_{m-r+1}^{(r)}-\sum_{k=r}^{m-1}g_k^{(r)}c_{k,m-r+1}^{(r-1)}}{c_{m,m-r+1}^{(r)}\prod^{m-1}_{k=r}\cos\varphi^{(r)}_k}
\end{align}
\end{subequations}
for $m\in\{r,r+1,\dots,N\}$. Notice that the right hand side of \eqref{eq:angles} only depends on $\theta_{m'}^{(r)}$ and $\varphi_{m'}^{(r)}$ with $m'< m$, and thus, we can solve the equations starting from $m=r$ towards $n$ iteratively to straightforwardly determine all of $\theta_m^{(r)}$ and $\varphi_m^{(r)}$.

For the first supermode, we have analytic solutions
\begin{align}
    \theta_{m}^{(1)}=\arg f_{m}^{(1)},~\varphi_{m}^{(1)}=\sin^{-1}\frac{\left|f_m^{(1)}\right|}{\sqrt{1-\sum_{k=1}^{m-1}\left|f_k^{(1)}\right|^2}},
\end{align}
while we generally need to employ numerical methods to demultiplex further supermodes. After solving for all $\theta_m^{(r)}$ and $\varphi_m^{(r)}$ and determining $\hat{R}^{(r)}$, it is straightforward to obtain $c_{\ell m}^{(r)}$ and confirm that \eqref{eq:lineartrans} is fulfilled by the above construction. The iterative procedure culminating in $\hat V = \hat V^{(s)}$ thus demultiplexes all $s$ supermodes into the leftmost $s$ spatial bins. We can then compute the reduced density matrix $\hat{\rho}_\mathcal{S}$ via \eqref{eq:matelements}.

\section{Quantum propagation of Kerr solitons}
\label{sec:soliton}
In this section, we study quantum propagation of a Kerr soliton. A classical soliton solution for \eqref{eq:nlse} with an average photon number of $\bar{n}$ takes a well-known sech form~\cite{Agrawal2019,Kivshar2003}
\begin{align}
    \label{eq:cont_sech}
    \phi_z^{(\mathrm{sech})}(t)=\frac{\bar{n}}{2}e^{\mathrm{i}\bar{n}^2t/8}\sech{\frac{\bar{n}z}{2}}.
\end{align}
While this solution is exact in the realm of classical optics where it may be thought of as specifying the pulse amplitudes of a coherent state, quantum mechanical effects such as squeezing~\cite{Haus1990,Carter1987}, quantum-induced dispersion of pulse envelope~\cite{Lai1989_b}, and soliton evaporation~\cite{villari2018} can become pronounced in the regime of large nonlinearity where $\bar n$ becomes small.

Conventionally, quantum mechanical soliton solitons can be treated using the time-dependent Hartree-Fock approximation (TDHF)~\cite{Haus1989}. As long as $\bar{n}\gg1$ holds true, TDHF predicts that a quantum soliton initialized as a coherent state of the envelope $\phi_z(0)$ approximately evolves according to~\cite{Wright1991}
\begin{align}
    \label{eq:tdhf}
    \ket{\Psi(t)}\approx e^{-\frac{\bar{n}}{2}}\sum_{n=0}\exp\left(\frac{\mathrm{i}(2n-\bar{n})\bar{n}nt}{8}\right)\frac{\alpha^n\hat{A}^{\dagger n}}{n!}\ket{0},
\end{align}
where $\alpha=\sqrt{\bar{n}}$, and $\hat{A}$ is the annihilation operator of the classical soliton-pulse supermode~\cite{Wright1991}. In discretized coordinates, 
$\hat{A}=\sum f^\text{(sech)}_m\hat{a}_m$, where $f^\text{(sech)}_m\propto\phi^{(\mathrm{sech})}_{m\Delta z-L/2}(0)$, up to a proportionality constant independent of $m$ such that $\boldsymbol{f}^\text{(sech)}$ is normalized. The main feature of the approximate TDHF solution is that it is closed within the subspace $\mathcal{S}$ of the soliton supermode $\boldsymbol{f}^\text{(sech)}$, so that, e.g., the reduced density matrix of $\ket{\Psi(t)}$ in the subspace $\mathcal{S}$ has unit purity throughout the dynamics of \eqref{eq:tdhf}. Nevertheless, due to the Kerr-type nonlinear phase shifts, \eqref{eq:tdhf} deviates from a coherent state as it evolves, leading to a variety of interesting phase-space dynamics~\cite{Korolkova2001,Wright1991,Singer1992,Yanagimoto2020}. On the other hand, it is difficult to quantify the accuracy or regime of validity of the TDHF due to its non-perturbative nature, and phase-space dynamics of quantum solitons beyond TDHF in the few-photon regime remain largely unexplored. In the following, we show that our MPS-based scheme serves as a powerful numerical tool to explore this latter regime.

\begin{figure*}[bt]
  \centering
    \includegraphics[width=0.98\textwidth]{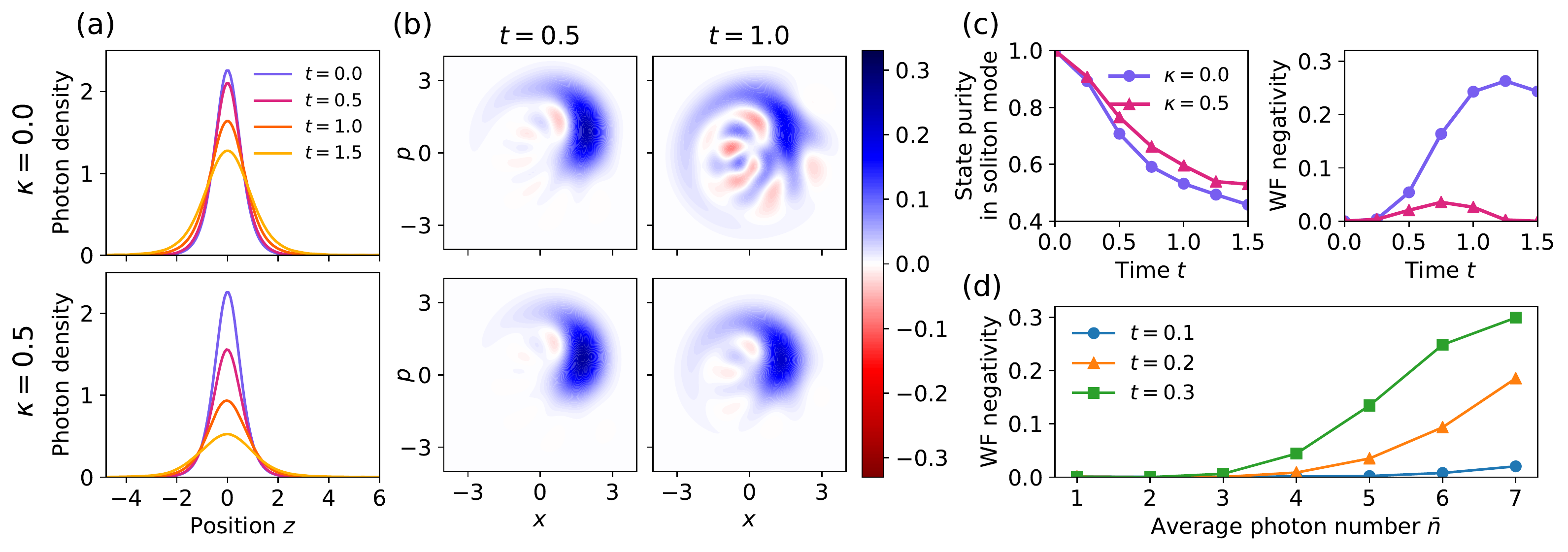}
    \caption{(a)--(c) Quantum simulations of a pulse instantiated as a coherent-state sech soliton using MPS with bond dimension $\chi=40$. Average photon number of $\bar{n}=3.0$ is used with ($\kappa=0.5$) and without ($\kappa=0$) linear loss. MCWF with $M=100$ quantum trajectories is used for the simulation with finite loss. (a) Time evolution of photon density distribution $\langle \hat{\phi}_z^\dagger\hat{\phi}_z\rangle$. (b) Wigner functions for the reduced density matrix $\hat{\rho}_\mathcal{S}$ for the sech supermode $\boldsymbol{f}^\text{sech}$. Upper and lower rows are for $\kappa=0.0$ and $\kappa=0.5$, respectively. (c) Time evolution of the purity $\mathrm{Tr}(\hat{\rho}_\mathcal{S}^2)$ (left figure) and the doubled volume of the Wigner function (WF) negativity~\cite{Kenfack2004} (lower figure) of the soliton pulse state. (d) WF negativity of the soliton pulse state at various times as a function of $\bar{n}$. MPS-based simulation with $\chi=50$ is used.}
    \label{fig:soliton}
\end{figure*}

To simulate the propagation of a soliton, we initialize a coherent-state MPS according to \eqref{eq:coherent} with envelope $\boldsymbol{f}^\text{(sech)}$ and with $\alpha=\sqrt{\bar{n}}$. Using TEBD, we simulate the time evolution of the state under the Hamiltonian \eqref{eq:bosehubbard} with and without linear loss, where the loss dynamics are simulated via MCWF with quantum jump operators $\{\sqrt{\kappa}\hat{a}_1,\dots,\sqrt{\kappa}\hat{a}_N\}$, where $\kappa$ is the power decay rate. Fig.~\ref{fig:soliton}(a) shows the time evolution of the photon density distribution (i.e., $g^{(1)}$ correlation function)  $\langle\hat{\phi}_z^\dagger\hat{\phi}_z\rangle$, for an initial pulse amplitude $\bar{n}=3$. We see that even on the level of $g^{(1)}$, the pulse envelope exhibits dispersion as a function of time~\cite{Lai1989_b}, reflecting the fact that the initial classical soliton is not an exact eigenstate of the quantum Hamiltonian; we sometimes refer to such non-classical dispersion as being ``quantum induced''.

We next consider the reduced density matrix $\hat\rho_\mathcal{S}$ for the sech-pulse supermode $\mathcal{S}$ using the demultiplexing scheme developed in Sec.~\ref{sec:demultiplexing} with the envelope function $\boldsymbol{f^\text{(sech)}}$. Consider first the case without loss, or $\kappa = 0$. Fig.~\ref{fig:soliton}(b) shows snapshots of the Wigner functions of $\hat\rho_\mathcal{S}$, which exhibit a substantial amount of Wigner function negativity and signify non-classicality in the state as might be expected for single-mode Kerr evolution. However, Fig.~\ref{fig:soliton}(c) shows that the purity $\mathrm{Tr}(\hat{\rho}_\mathcal{S}^2)$ exhibits a monotonic decay in time, indicating that the sech-pulse subspace $\mathcal{S}$ is not closed under the dynamics and in fact, coupling between $\mathcal{S}$ and the rest of the system $\mathcal{E}$ can act as an effective decoherence channel. Actually, due to the nonlinear nature of the dynamics, $\hat{\rho}_\mathcal{S}$ may not be pure for any choice of a supermode $\boldsymbol{f}$ in general. Also shown in Fig.~\ref{fig:soliton}(c) is the volume of the Wigner function negativity, which serves as a measure of the non-classicality of the state~\cite{Kenfack2004}. Following an initial increase, the volume of Wigner function negativity starts to decrease after some time, again due to competition between the nonlinear dynamics and the effective decoherence caused by entanglement with $\mathcal{E}$. These features are in stark contrast to the single-mode dynamics predicted by TDHF.

We can also contrast this effective decoherence due to entanglement between $\mathcal{S}$ and $\mathcal{E}$ with standard dissipation due to linear loss. When loss is incorporated to the simulation, we obtain an ensemble of quantum trajectories $\{\ket{\Psi_1(t)},\dots,\ket{\Psi_{M}(t)}\}$ via the MCWF method~\cite{Wiseman2009}. The final reduced density matrix is calculated by averaging the reduced density matrices over the ensemble according to $\hat{\rho}_\mathcal{S}=M^{-1}\sum_{i=1}^{M}\hat{\rho}_{\mathcal{S},i}$, where $\hat{\rho}_{\mathcal{S},i}$ is the reduced density matrix of $\ket{\Psi_i}$. As expected, the linear loss causes a decay in the amplitude of the photon density distribution as shown in Fig.~\ref{fig:soliton}(a), while quantum mechanically, Fig.~\ref{fig:soliton}(b) and (c) show that the non-classical features of the state are critically diminished by the presence of the linear loss.

We additionally investigate how the amount of excitation affects the phase-space dynamics of pulse propagation. Classically, due to the nonlinear nature of the interactions, the effective nonlinear rate is expected to be enhanced when the peak pulse intensity is larger. In Fig.~\ref{fig:soliton}(d), we show the volume of the Wigner function negativity as a function of the average photon number $\bar{n}$ in the soliton pulse, where we observe that larger pulse excitations increase the rate at which non-classical features are formed, confirming the classical intuition in the few-photon regime.

\begin{figure}[hbt]
    \centering
      \includegraphics[width=0.48\textwidth]{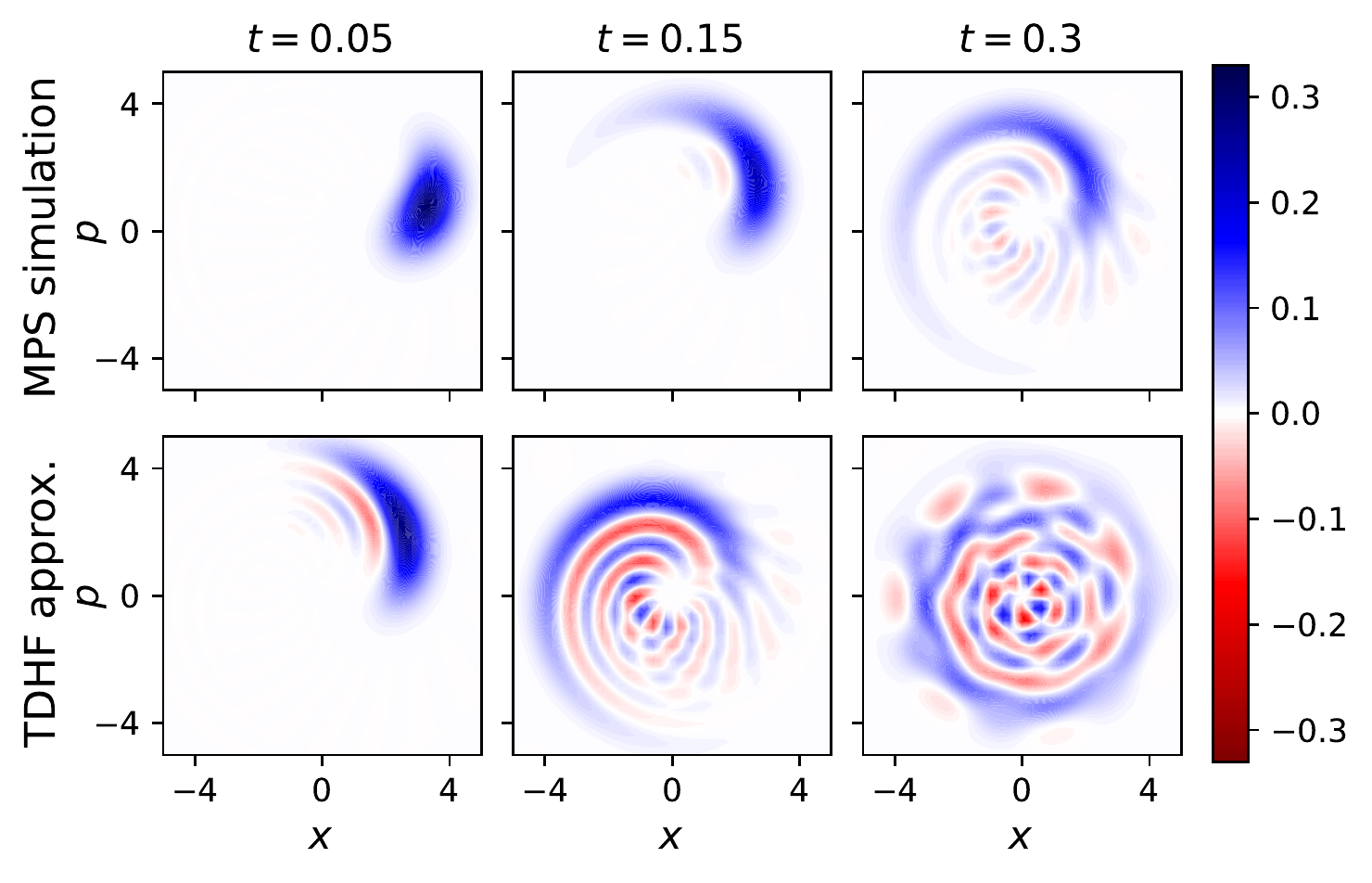}
      \caption{Wigner functions of soliton pulse state with $\bar{n}=6$ obtained using MPS-based full-quantum simulation with $\chi=50$ (upper row) and the time-dependent Hartree-Fock approximation (TDHF) following \eqref{eq:tdhf} (lower row).} 
 \label{fig:tdhf}
  \end{figure}

Finally, to highlight the difference between full numerical simulation and TDHF, Fig.~\ref{fig:tdhf} compares the Wigner functions of a soliton pulse initialized with $\bar{n}=6$ obtained by the MPS-based simulation against TDHF \eqref{eq:tdhf}. While they exhibit qualitatively similar interference patterns, the discrepancies in the time at which a similar phase-space structure is reached indicate that TDHF is overestimating the rate of the nonlinear phase-shift (see $t=0.3$ for MPS-based simulation and $t=0.15$ for TDHF, for instance). Additionally, full quantum simulation results exhibit visible reduction in the magnitude of Wigner function negativity compared to TDHF, highlighting the effects of aforementioned decoherence due to entanglement between $\mathcal{S}$ and $\mathcal{E}$. These discrepancies point to the presence of broadband physics beyond TDHF in soliton propagation.

\begin{figure}[tb]
    \includegraphics[width=0.5\textwidth]{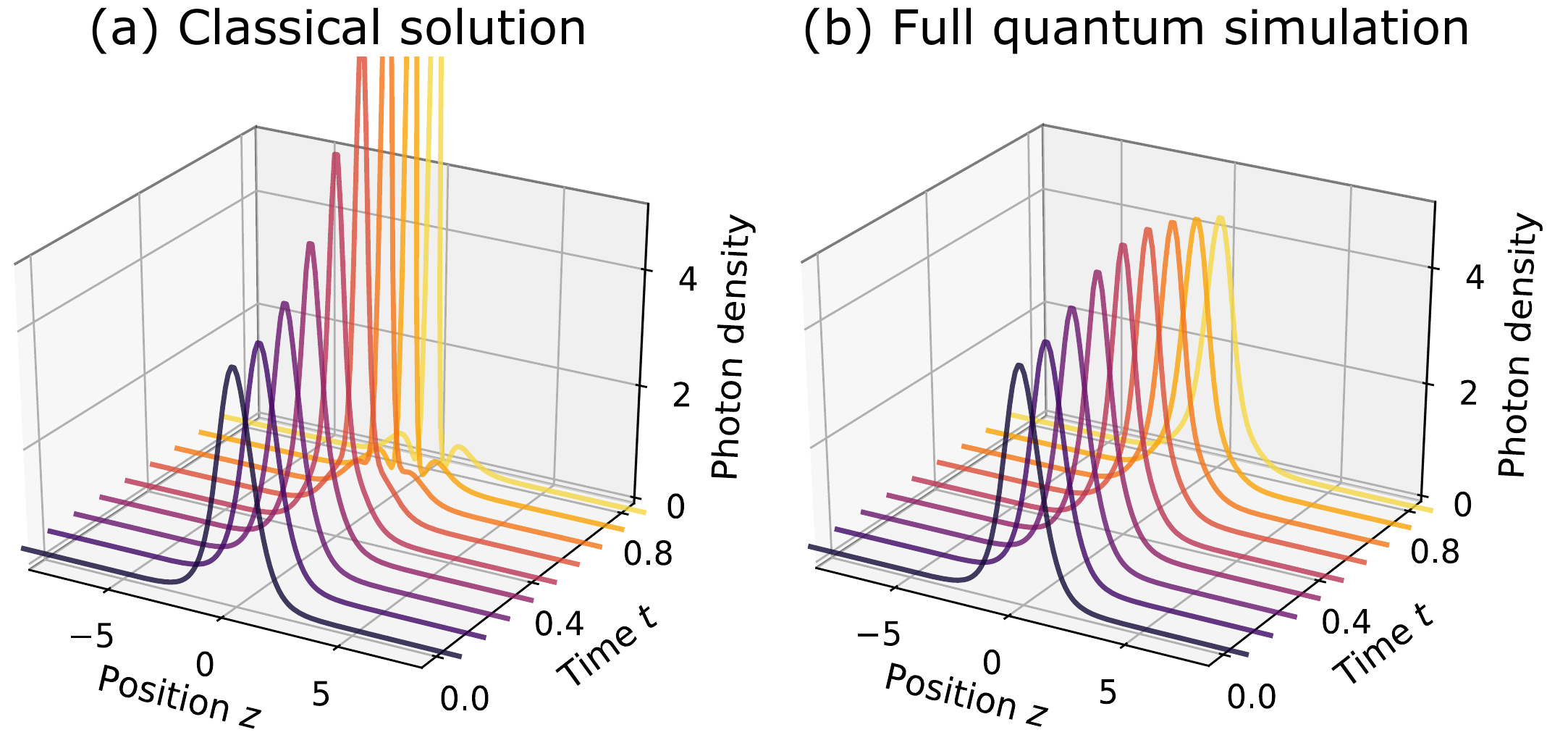}
    \caption{Time evolution of the photon density distribution of a pulse instantiated as a coherent-state second-order soliton with mean photon number $\bar{n}^{\text{(2sech)}}=8$. (a) Classically expected dynamics given by \eqref{eq:soliton2}. (b) Full quantum simulation under the Hamiltonian \eqref{eq:liebliniger} using MPS with bond dimension $\chi=60$.}
    \label{fig:soliton2}
\end{figure}

While the quasi-stationary nature of the quantum dynamics of fundamental solitons is in qualitative agreement with their classical behavior, it is in fact possible for highly-quantum photon dynamics to exhibit much more striking deviations from classical behavior. This is the case, for example, if we apply our simulation method to study the quantum propagation of a second-order soliton with classical waveform~\cite{Kivshar2003}
\begin{align}
    \label{eq:soliton2}
   \phi_z^{(\mathrm{2sech})}(t)=\frac{2e^{\mathrm{i}\bar{n}^2t/8}\bar{n}\left(3e^{\mathrm{i}\bar{n}^2t}\cosh(\bar{n}z/2)+\cosh(3\bar{n}z/2)\right)}{3\cos(\bar{n}^2t)+4\cosh(\bar{n}z)+\cosh(2\bar{n}z)},
\end{align}
which, classically, is a periodic ``breather'' solution of the NLSE \eqref{eq:nlse}. The average photon number of the second-order soliton is $\bar{n}^{\text{(2sech)}}=4\bar{n}$, where $\bar{n}$ is the average photon number in its corresponding fundamental soliton. At $t=0$, this means the waveform field amplitude of the second-order soliton is twice that of the fundamental soliton, i.e., $\phi_z^{(\mathrm{2sech})}(0)=2\phi_z^{(\mathrm{sech})}(0)$. After $t=0$, as shown in Fig.~\ref{fig:soliton2}(a), the classical waveform of the second-order soliton exhibits significant narrowing and a characteristic triplet structure. On the other hand, as shown in Fig.~\ref{fig:soliton2}(b), full quantum evolution of a second-order soliton instantiated in a few-photon coherent state of \eqref{eq:soliton2} at $t = 0$ exhibits qualitatively different dynamics. While the photon density distribution exhibits some narrowing of the pulse width initially, the peak pulse intensity fails to reach the level expected from the classical solution. Moreover, no signature of the triplet structure is observed. We attribute these features to the quantum-induced dispersion of the pulse envelope that we also observed in the quantum propagation of fundamental solitons~\cite{Lai1989_b}, which appears to play a more critical role in the evolution of this higher-order soliton.

\section{Quantum propagation of Simultons}
\label{sec:simulton}
In this section, we apply our technique to a 1D $\chi^{(2)}$ nonlinear waveguide in which interactions occur between a fundamental harmonic (FH) band and a second harmonic (SH) band. After normalization with respect to time and space, the system Hamiltonian takes the form~\cite{Drummond1997,Raymer1991}
\begin{align}
    \label{eq:chi2hamiltonian}
    \begin{split}
    \hat{H}=&-\frac{1}{2}\int\mathrm{d}z\left(\hat{\phi}_z^\dagger\partial_z^2\hat{\phi}_z+\beta\hat{\psi}_z^\dagger\partial_z^2\hat{\psi}_z\right)\\
    &\qquad+\frac{1}{2}\int\mathrm{d}z\left(\hat{\phi}_z^{\dagger 2}\hat{\psi}_z+\hat{\phi}_z^2\hat{\psi}_z^\dagger\right),
    \end{split}
\end{align}
where $\hat{\phi}_z$ and $\hat{\psi}_z$ are respectively FH and SH local field annihilation operators with commutation relationships $[\hat{\phi}_z,\hat{\phi}^\dagger_{z'}]=[\hat{\psi}_z,\hat{\psi}^\dagger_{z'}]=\delta(z-z')$. We have assumed that the FH and SH carriers are group-velocity matched, while $\beta$ represents the group velocity dispersion of SH relative to FH. \eqref{eq:chi2hamiltonian} can be discretized in space to give
\begin{subequations}
\label{eq:chi2}
\begin{align}
    \hat{H}=\sum_{m}\left(\hat{H}_{\text{a},m}+\hat{H}_{\text{b},m}+\hat{H}_{\text{NL},m}\right),
\end{align}
with
    \begin{align}
        &\hat{H}_{\text{a},m}=-\frac{1}{2\Delta z^2}\left(\hat{a}_{m+1}^\dagger\hat{a}_m+\hat{a}_m\hat{a}_{m+1}^\dagger-2\hat{a}_m^\dagger\hat{a}_m\right)\\
        &\hat{H}_{\text{b},m}=-\frac{\beta}{2\Delta z^2}\left(\hat{b}_{m+1}^\dagger\hat{b}_m+\hat{b}_m\hat{b}_{m+1}^\dagger-2\hat{b}_m^\dagger\hat{b}_m\right)\\
        &\hat{H}_{\text{NL},m}=\frac{1}{2\sqrt{\Delta z}}\left(\hat{a}_m^{\dagger 2}\hat{b}_m+\hat{a}_m^{2}\hat{b}_m^\dagger\right),
    \end{align}
\end{subequations}
where $\hat{a}_m$ and $\hat{b}_m$ are the FH and SH field annihilation operators for the $m$th spatial bin, respectively.
\begin{figure}[t!]
  \centering
    \includegraphics[width=0.43\textwidth]{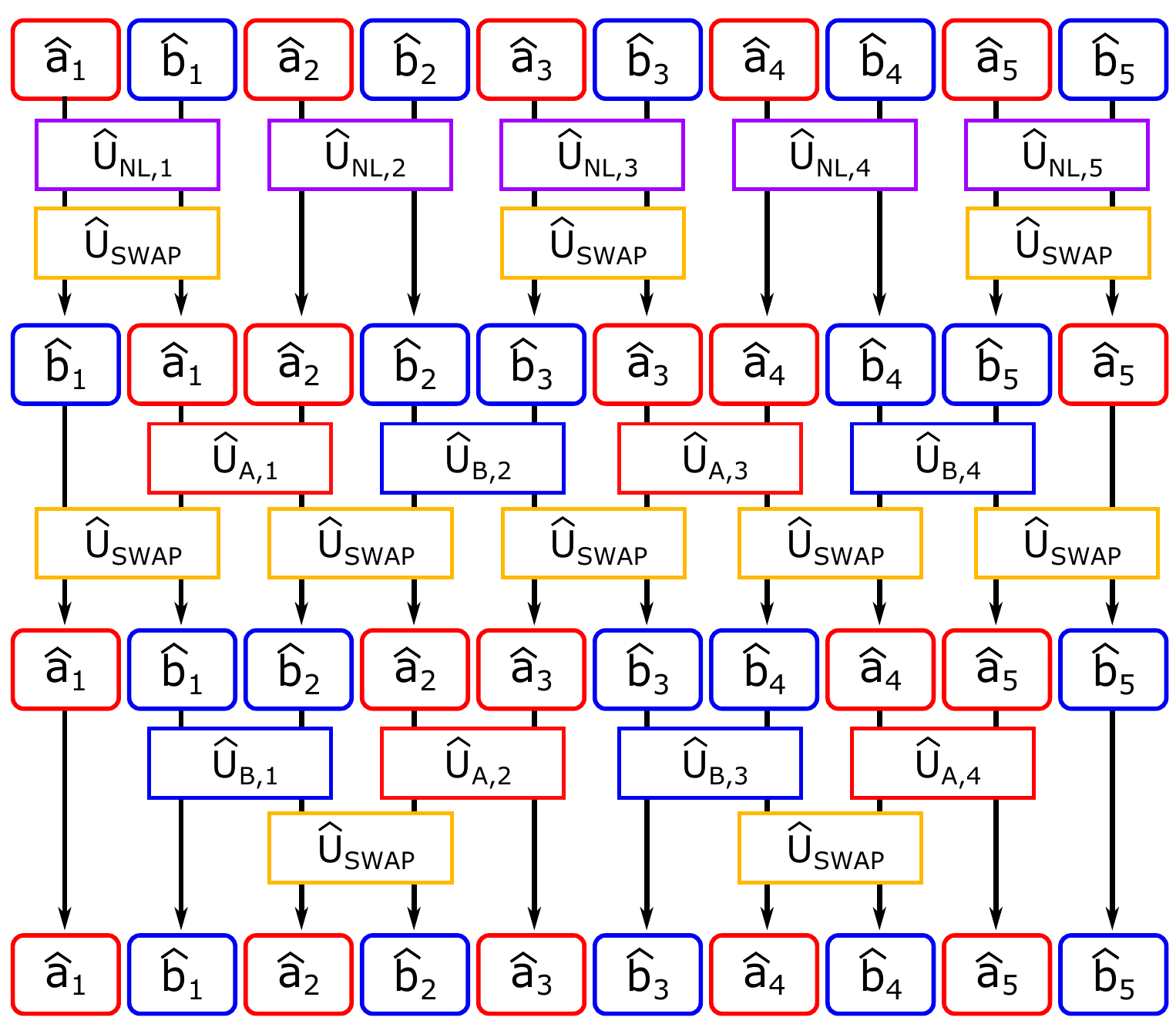}
    \caption{An example implementation of a short-time evolution under the $\chi^{(2)}$ Hamiltonian \eqref{eq:chi2hamiltonian} using TEBD. Discretized FH modes $\{\hat{a}_1,\dots,\hat{a}_N\}$ and SH modes $\{\hat{b}_1,\dots,\hat{b}_N\}$ are encoded on an MPS in an alternating manner. Swap operations $\hat{U}_\text{SWAP}$ are used to bring distant modes together. Two-mode operations $\hat{U}_{\mathrm{a},m}$, $\hat{U}_{\mathrm{b},m}$, and $\hat{U}_{\mathrm{NL},m}$ are for FH dispersion, SH dispersion, and nonlinear three-wave mixing interaction, respectively.}
    \label{fig:chi2}
\end{figure}

The presence of both FH and SH fields requires some care in applying the two-mode operations for implementing TEBD for the Hamiltonian \eqref{eq:chi2}. One approach is shown schematically in Fig.~\ref{fig:chi2}, where we prepare $2N$ MPS bins to represent $N$ spatial bins, with FH and SH modes encoded in an alternating manner. To apply Trotterization as in Sec.~\ref{sec:mps}, a short-time unitary evolution $e^{-\mathrm{i}\hat{H}\delta t}$ is decomposed into two-mode operations $\hat{U}_{\lambda,m}=e^{-\mathrm{i}\hat{H}_{\lambda,m}\delta t}~(\lambda=\text{a},\text{b},\text{NL})$ and applied as shown in Fig.~\ref{fig:chi2}. Since $\hat{a}_m$ and $\hat{a}_{m+1}$ are no longer next to each other in this representation, we also utilize a swap operation $\hat{U}_\text{SWAP}$~\cite{Vidal2003} to bring these modes together in an alternating manner for the application of $\hat{U}_{\mathrm{a},m}$ (and similarly for $\hat{U}_{\mathrm{b},m}$).

\begin{figure}[tb]
  \centering
    \includegraphics[width=0.5\textwidth]{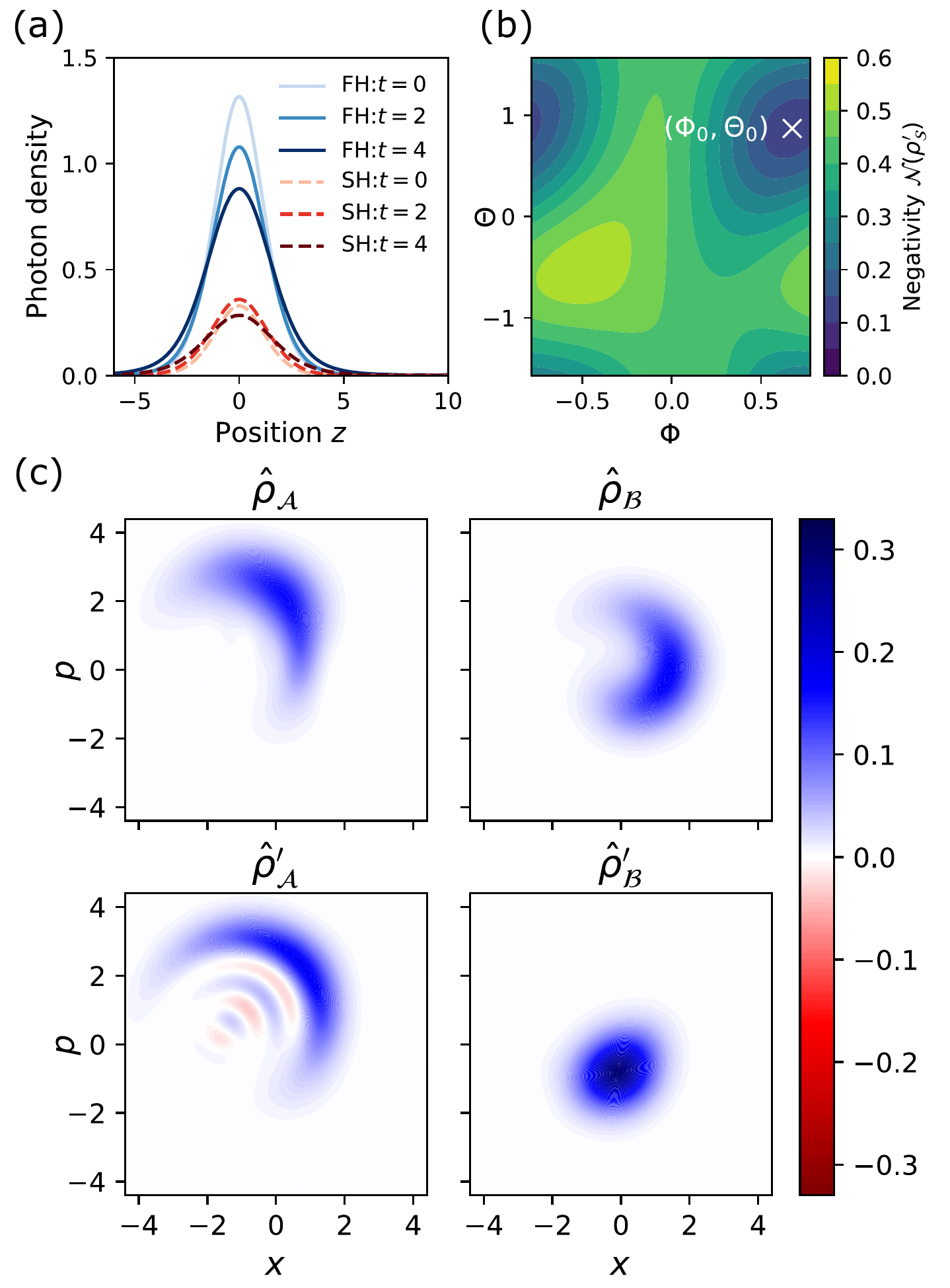}
    \caption{Quantum simulations of pulses instantiated as a coherent-state simulton with FH photon number $\bar{n}=4$. MPS with bond dimension $\chi=50$ is used. (a) Time evolution of the FH and SH photon density distribution. (b) Negativity $\mathcal{N}(\rho_\mathcal{S}')$ calculated for the two-mode reduced density matrix under linear transformations $\hat{\rho}_\mathcal{S}'=\hat{W}^\dagger\hat{\rho}_\mathcal{S}\hat{W}$, where $\hat{W}$ is parametrized by a pair of angles $(\Phi,\Theta)$. White cross indicates  $(\Phi_0,\Theta_0)\approx(0.21\pi,0.28\pi)$ that minimizes $\mathcal{N}(\hat{\rho}_\mathcal{S}')$ for $t=4$. (c) Wigner function of the single-mode reduced density matrix before ($\hat{\rho}_\mathcal{A,B}$) and after ($\hat{\rho}_\mathcal{A,B}'$) applying $\hat{W}(\Phi_0,\Theta_0)$ at $t=4$. Labels $\mathcal{A}$ and $\mathcal{B}$ represent FH and SH, respectively.}
    \label{fig:chi2_combined}
\end{figure}

The Hamiltonian \eqref{eq:chi2hamiltonian} supports various classical soliton solutions~\cite{Buryak1995,Buryak2002}. Specifically, an analytic ``simulton'' solution exists for $\beta=2$ with the form~\cite{Werner1993}
\begin{subequations}
    \label{eq:simulton_waveform}
\begin{align}
    \phi_z(t)&=\phi_0\mathrm{sech}^2\left(\sqrt{\phi_0/6}~z\right)e^{\mathrm{i}\phi_0 t/3}\\
    \psi_z(t)&=-\frac{\phi_0}{2}\mathrm{sech}^2\left(\sqrt{\phi_0/6}~z\right)e^{2\mathrm{i}\phi_0 t/3},
\end{align}
\end{subequations}
where $\phi_0$ is related to the average FH photon number $\bar{n}$ via $\phi_0=\sqrt[3]{3\bar{n}^2/32}$. As was done for \eqref{eq:cont_sech}, we discretize and normalize \eqref{eq:simulton_waveform} at $t=0$ to construct FH and SH supermodes with annhilation operators denoted $\hat{A}$ and $\hat{B}$, respectively. The initial simulton MPS state is that of a coherent state in $\hat{A}$ and $\hat{B}$ with displacements $\sqrt{\bar{n}}$ and $-\sqrt{\bar{n}}/2$, respectively. In Fig.~\ref{fig:chi2_combined}(a), we show the time evolution via TEBD of the photon density distribution of a pulse initialized in the classical simulton supermode. As for the case of the Kerr soliton, the dynamics show a quantum-induced dispersion of the pulse envelope not predicted by classical dynamics; as part of this process, we also numerically observe a slight exchange of excitations between FH and SH.

To investigate the joint phase-space properties of the FH and SH pulse supermodes, we calculate two-mode reduced density matrix $\hat{\rho}_\mathcal{S}$ where $\mathcal{S}=\mathcal{A}\otimes\mathcal{B}$, the joint Hilbert space of FH and SH, respectively. In general, the state of the system features entanglement between $\mathcal A$ and $\mathcal B$; to quantify this entanglement, we utilize the negativity measure $\mathcal{N}(\hat{\rho})$ (not to be confused with the Wigner function negativity used earlier)~\cite{Vidal2002}, defined for a bipartite density matrix $\hat\rho$ as
\begin{align}
    \mathcal{N}(\hat{\rho})=\textstyle\frac12(1-\Vert\hat{\rho}^{\mathrm{T}_\mathcal{A}}\Vert_1),
\end{align}
where $\mathrm{T}_\mathcal{A}$ is partial transposition with respect to the first mode, and $\norm{\cdot}_1$ is the trace norm. Here, $\mathcal{N}(\hat{\rho})>0$ serves as a sufficient condition for the entanglement.

Generally, we can change the value of the $\mathcal{N}$ by considering hybrid mixtures of the modes $\mathcal A$ and $\mathcal B$. Finding the linear combination that best ``disentangles'' $\mathcal A$ and $\mathcal B$ therefore provides insight into their entanglement structure. More specifically, we introduce a two-mode linear operation $\hat{W}(\Phi,\Theta)=\exp\left\{\Phi\left(e^{\mathrm{i}\Theta}\hat{A}^\dagger\hat{B}-e^{-\mathrm{i}\Theta}\hat{A}\hat{B}^\dagger\right)\right\}$ with $-\pi/4\leq \Phi\leq \pi/4$ and $-\pi/2\leq \Theta\leq \pi/2$, which implements the transformation $\hat{\rho}_\mathcal{S}'=W^\dagger\hat{\rho}_\mathcal{S}\hat{W}$. Importantly, the transformation $\hat{W}_0=\hat{W}(\Phi_0,\Theta_0)$ that minimizes $\mathcal N(\hat\rho_{\mathcal{S}}')$ is expected to maximize the amount of information available in the single-mode reduced density matrices $\hat{\rho}_\mathcal{A}'=\mathrm{Tr}_\mathcal{B}(\hat{\rho}_\mathcal{S}')$ and $\hat{\rho}_\mathcal{B}'=\mathrm{Tr}_\mathcal{A}(\hat{\rho}_\mathcal{S}')$.

In Fig.~\ref{fig:chi2_combined}(b), for the final state of the pulse propagation at $t=4$, we map the negativity $\mathcal{N}(\hat{\rho}_\mathcal{S}')$ for various $(\Phi,\Theta)$, which shows a clear minimum at the marked position of $(\Phi_0,\Theta_0)$. In Fig.~\ref{fig:chi2_combined}(c), we show the Wigner functions of the single-mode reduced density matrices before and after the transformation $\hat{W}_0$. Before applying the transformation, the Wigner functions of both $\hat{\rho}_\mathcal{A}$ and $\hat{\rho}_\mathcal{B}$ are crescent-shaped with no negativity, indicating that they are highly mixed states due to the entanglement between FH and SH. On the other hand, after application of $\hat{W}_0$, the Wigner function of $\hat{\rho}_\mathcal{A}'$ remarkably exhibits considerable non-classicality, while the Wigner function of $\hat{\rho}_\mathcal{B}'$ resembles that of a coherent state. This reveals a somewhat surprising feature of the simulton quantum dynamics: a hybrid supermode $\hat{A}_0=\cos\Phi_0\hat{A}+e^{\mathrm{i}\Theta_0}\sin\Phi_0\hat{B}$, composed of both FH and SH components, is the one which predominantly experiences strongly nonlinear dynamics, while the other hybrid supermode $\hat{B}_0=\cos\Phi_0\hat{B}-e^{-\mathrm{i}\Theta_0}\sin\Phi_0\hat{A}$ experiences little nonlinearity. A similar analysis can, in principle, be applied to a broader class of solitons and general pulse propagation~\cite{Kivshar2003}.

\section{Conclusion}
\label{sec:conclusion}
In this research, we have motivated the use of MPS techniques to efficiently represent and simulate a quantum optical pulse as it dynamically propagates through a nonlinear 1D waveguide. In doing so, we have developed a numerical method to overcome the problem of efficiently accessing and manipulating nonlocal pulse supermodes of the local MPS representation, allowing us to view for the first time the full quantum dynamics of the pulse in a phase-space picture. As a demonstration, we have performed quantum simulations of Kerr soliton propagation and observed that the phase-space portraits of an initially classical sech-pulse supermode can evolve highly non-classical features, i.e., Wigner function negativity. These results have been contrasted with predictions based on TDHF, highlighting the presence of rich quantum dynamics beyond conventional approximations for quantum Kerr solitons. We have also extended our analysis to the quantum propagation of a $\chi^{(2)}$ simulton and have revealed unexpected entanglement structure between the FH and SH pulses of the simulton, identifying a hybrid supermode that predominantly exhibits non-classical features. Our scheme is compatible with local dissipation associated with, e.g., waveguide losses, and, more generally, could be applied to any one-dimensional photonic system in principle. Considering the rapid recent progress towards single-photon nonlinearities in dispersion-engineered and highly nonlinear nanophotonic platforms, it is of imminent interest to establish a unified theoretical framework in which to understand the quantum dynamics of photons in such devices. To this end, our work takes a step towards bridging the significant conceptual gaps between classical wave optics, CV photonic quantum information, and strongly-interacting quantum many-body physics, all of which are expected to play important roles in conceptualizing and engineering the future of broadband quantum optics.
\section*{Funding}
Army Research Office (W911NF-16-1-0086); National Science Foundation (CCF-1918549, PHY-2011363).
\section*{Acknowledgments}
R.\,Y.\ developed the numerical techniques, performed the simulations, and generated the figures. E.\,N.\ and H.\,M.\ advised and directed the project. R.\,Y.\ and E.\,N.\ wrote the manuscript with detailed input and feedback from all authors. All authors contributed significantly to the conception of the project.

The authors wish to thank NTT Research for their financial and technical support. R.\,Y.\ would like to thank Tomohiro Soejima for helpful discussions. R.\,Y.\ is supported by Stanford Q-FARM Ph.D.\ Fellowship and Masason Foundation. 
\section*{Disclosures}
The authors declare no conflicts of interest.

\bibliography{myfile}
\end{document}